\documentclass[twocolumn,amsmath,amssymb]{revtex4}
\usepackage{graphicx} % Include figure files
\usepackage{subfigure}
\usepackage{dcolumn} % Align table columns on decimal point
\usepackage{bm} % bold math
\usepackage{color}
\usepackage{ulem}
\usepackage{subfigure}
\usepackage{float}

\newcommand{\rev}[1]{{\color{black}{#1}}}
\newcommand{\revdos}[1]{{\color{black}{#1}}}
\newcommand{\revtres}[1]{{\color{black}{#1}}}
\begin{document}

\title{Resistive thrust production can be as crucial as added mass mechanisms\\ for inertial undulatory swimmers}

\author{M. Pi\~neirua }
\author{R. Godoy-Diana}
\author{B. Thiria}%\email{bthiria@pmmh.espci.fr}

\affiliation{Laboratoire de Physique et M\'ecanique des Milieux H\'et\'erog\`enes, CNRS, ESPCI ParisTech, UPMC Paris 6, Universit\'e Paris Diderot, 10 rue Vauquelin, 75005 Paris, France.}

\begin{abstract}
{In this paper, we address a crucial point regarding the description of moderate to high Reynolds numbers aquatic swimmers. For decades, swimming animals have been classified in two different families of propulsive mechanisms based on the Reynolds number: the “resistive” swimmers, using local friction to produce the necessary thrust force for locomotion at low Reynolds number and the “reactive” swimmers, lying in the high Reynolds range, and using added mass acceleration (described by perfect fluid theory). However, inertial swimmers are also systems that dissipate energy, due to their finite size, therefore involving strong resistive contributions, even for high Reynolds numbers. Using a complete model for the hydrodynamic forces, involving both reactive and resistive contributions, we revisit here the physical mechanisms responsible for the thrust production of such swimmers. We show, for instance, that the resistive part of the force balance is as crucial as added mass effects in the modeling of the thrust force, especially for elongated species.
The conclusions brought by this work may have significant contributions to the understanding of complex swimming mechanisms, especially for the future design of artificial swimmers.
}
\end{abstract}

\maketitle

Every fluid dynamicist has opened, at least once, a book addressing the mechanics of swimming. Although the problem has been studied by experimental biologists for almost a century \cite{Gray:1933}, and formalised later by the pioneer works of Taylor \cite{Taylor:1952} and Lighthill \cite{Lighthill:1969}, it remains a very active field for experimental and theoretical physics and biology (see e.g. recent reviews by \cite{Triantafyllou:2000,Liao:2007,Wu:2011}). Behind the elegant undulatory kinematics that leads to motion, it is Newton's third law that allows the estimation of the net thrust force produced by the animal. Basically, the local force applied by the fluid to the body in reaction to the body movements has two components: a \textit{resistive} component due to local friction at the fluid/solid interface and a \textit{reactive} (inertial) component coming from the amount of fluid accelerated away from the swimmer's body. The presence of these two contributions has brought scientists to make a distinction between different swimming mechanisms, depending on the animal body size or the nature of the fluid. For instance, swimmers at small scales are in the low-Reynolds domain where viscosity prevails over inertial effects. \rev{The swimming theory associated to those regimes is thus only based on local friction and is referred to as \emph{resistive theory} \cite{Hancock:1953,Gray:1955}. On the other hand, Lighthill \cite{Lighthill:1960,Lighthill:1971} and Wu \cite{Wu:1961}, }established a potential flow theory for inertial swimmers (high Reynolds domain) where viscous contributions are neglected, relying on a slender-body approximation that allows to integrate the \emph{reactive} lateral forces along the coordinate following the spinal cord of the fish. 

\begin{figure}[htb!]
\includegraphics[width=0.85\linewidth]{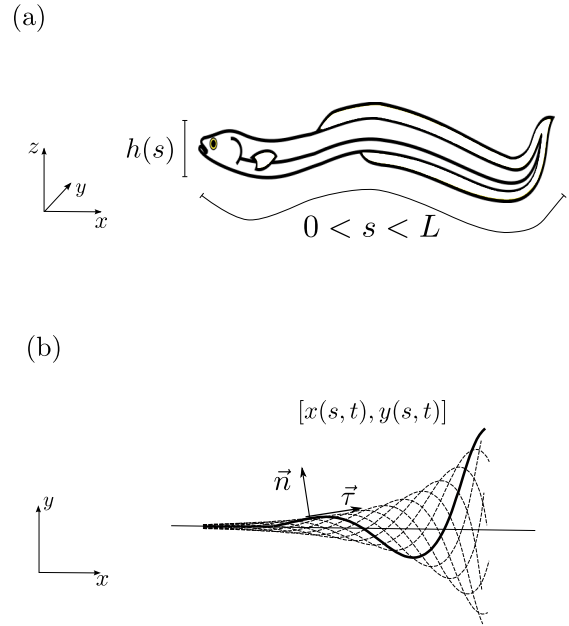}
\caption{Schematic diagrams of (a) an elongated fish of length $L$  where the function $h(s)$ describes its span varying along the longitudinal coordinate, and (b) the $y(s)$ function describing the undulation of the spinal cord of the fish.\label{fig_schema_fish}}
\end{figure}

The existence of these two models has led to a virtual frontier between two groups of swimmers in terms of the Reynolds number: the first being based on dissipation (small scale swimmers rely on the anisotropy of the friction drag components normal and tangential to each body section), and the second, used by large swimmers, based on inertial momentum transfer. In other words, thrust production would be based in the first case on local transversal \textit{velocities}, and in the second case, on local transversal \textit{accelerations}. At low Reynolds numbers, the locomotion problem is fully solved once the expression for the local drag is integrated. On the contrary, in the inertial regime, an additional model for the global drag experienced by the swimmer is needed to close the locomotion problem \rev{(see e.g. \cite{Alben:2012,Raspa:2014,Gazzola:2014,Gazzola:2015}). However recent works have shown that,  in order to give an accurate description of real swimmers \cite{Eloy:2013,Ramananarivo:2013,Porez:2014},  the local  balance of forces   normal to the body section needs an extra term accounting for the local dissipation due to lateral body motion. This term is referred to as ``quadratic drag" and expresses the effect of viscosity for these inertial regimes  \cite{Taylor:1952}, which determines the large flow separations occurring on finite size geometries \cite{Lian:1989} such as those involved in animal swimming. This \textit{resistive} contribution is a form drag that depends on the local velocity, and has a component in the swimming direction that can produce thrust. }

In the present work we propose to use such a local model \rev{for the normal forces}, involving both reactive and resistive contributions, to revisit the crucial question of the physical mechanism responsible for thrust production in moderate to high Reynolds number fish-like swimmers. \rev{The problem is posed in a general form as done by Eloy \cite{Eloy:2013}, for instance. However, here we use real fish kinematics from the literature to close the locomotion problem, thus avoiding the introduction of a skin friction model.} The swimmers are characterized through their geometrical aspect ratio $AR$ defined  as $AR=\max(h(s))/L$, with $L$ the total length of the swimmer and $h(0<s<L)$ the local height as a function of the curvilinear coordinate $s$ (see Fig. \ref{fig_schema_fish}-a). The latter will be considered as a constant $h(s)=H$ for the following analysis, where the swimmers are  modeled by infinitely thin rectangular foils. The swimming kinematics is characterized by the deformation of the spinal cord whose local position can be described by the $x(s,t)\,,y(s,t)$ coordinates, dependent of the curvilinear coordinate $s$ and time $t$ (see Fig.  \ref{fig_schema_fish}-b). During the imposed swimming motion, each slice of the swimmer is subjected to local forces corresponding to both reactive and resistive contributions.  Considering the inextensibility of the spinal cord, and using a second-order non-linear approach as in  \cite{Ramananarivo:2013,Eloy:2013,Eloy:2012}, the reactive and resistive forces per unit surface can be written as :
 \rev{ 
\begin{eqnarray}
\mathbf{f}_{ma}&=&-\mathcal{M}(h)(\ddot{y}+2U\dot{y'}+U^2y'')\mathbf{n} \; , \label{Fma}\\
\mathbf{f}_d&=&-\frac{1}{2} \rho  C_d \vert \dot{y} +U y' \vert (\dot{y} +U y' )\mathbf{n}  \;, 
\label{Fd}
\end{eqnarray} 
}
\noindent where $\mathcal{M}(h)$ represents the local added mass accelerated during swimming, $\mathbf{n}$ and $\mathbf{t}$ are the unity vectors normal and tangential to the fish surface respectively (see Fig. \ref{fig_schema_fish}-b) and the dot and prime symbols are time and space derivatives, respectively. In addition, $\rho$ is the fluid density and $C_d$ is a drag coefficient weighing the \emph{non-linear resistive term}. $C_d$ is associated to the  dynamic stalls at each swimming cycle that result from the large transversal local velocities and the finite geometry of the fish section. \rev{The lateral Reynolds numbers involved in the cases studied in this work range from 2000 to 30000, for which a constant value of  $C_d\sim2$ can be accurately used \cite {White:1998,Eloy:2013}.}  As evoked above, the consideration of the resistive component to accompany the classical potential flow model is a major point for the description of fish swimming mechanics. This point will be the core of the forthcoming discussion in this work, where we examine the role of both the acceleration of added mass and the quadratic hydrodynamic resistance in the production of thrust. The projection of Eqs. \ref{Fma} and \ref{Fd} in the swimming direction (in this case $\mathbf{-e_x}$) gives the contribution of these forces to the thrust. They read, respectively: 
\revdos{
\begin{eqnarray}
t_{ma}&=&-\mathcal{M}(h)(\ddot{y}+2U\dot{y'}+U^2y'')y' \; ,  \label{Fmax}\\
t_d&=&-\frac{1}{2} \rho  C_d \vert \dot{y} +U y' \vert (\dot{y} +U y' )y' \; . \label{Fdx}
\end{eqnarray} 
 }
In order to compare these two terms for a given swimmer (defined by its kinematics and aspect ratio),  the knowledge of both $C_d$ and $\mathcal{M}(h)$ are needed. We have estimated the precise value of  the added mass coefficient $\mathcal{M}(h)$ for the present rectangular foils by studying the impulse response of elastic plates of different aspect ratios in water.  $\mathcal{M}(h)$ is then deduced through the modification of the relaxation frequency of the plate, which changes with the fluid loading \revtres{(see Appendix \ref{app_addmass})}.  For slender body swimmers ($AR<0.4$), we have confirmed the linear dependence of $\mathcal{M}(h)$  on the aspect ratio, as reported in previous works \cite{Eloy:2012, Eloy:2013}.  Thus, we shall consider the added mass coefficient as

\begin{equation}
\mathcal{M}(h)=\frac{\pi}{4}\rho h.
\end{equation}  

\begin{figure}[t]
\begin {tabular} {l @{} l}
(a)&(b)\\
\includegraphics[height=0.4\linewidth]{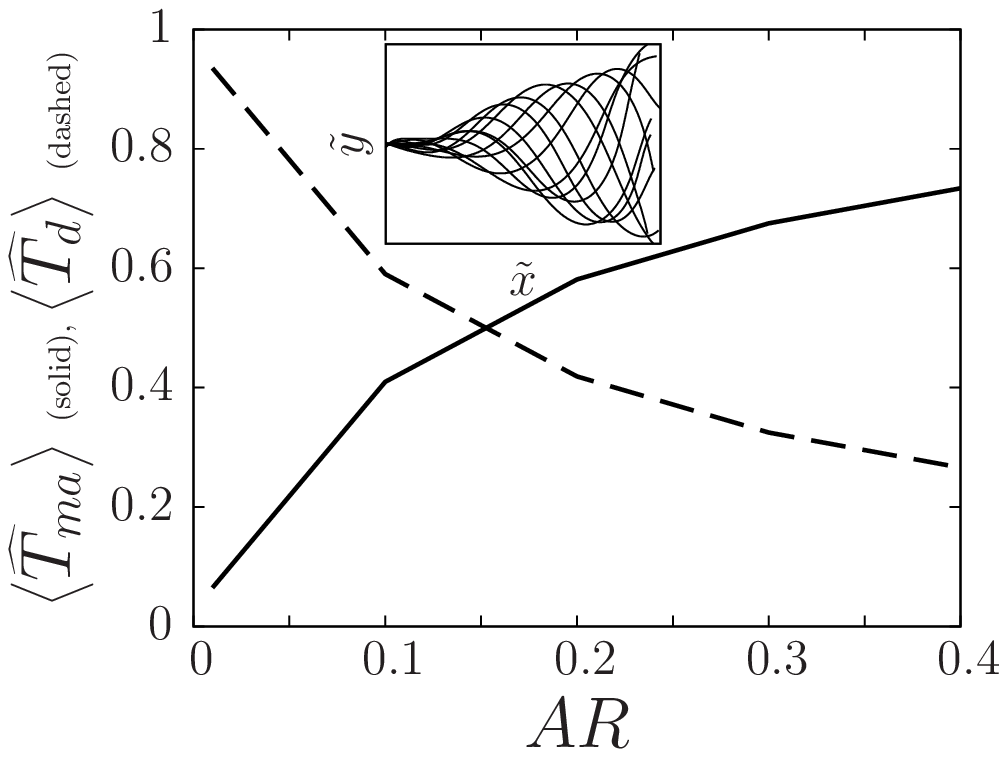}&
\includegraphics[height=0.4\linewidth]{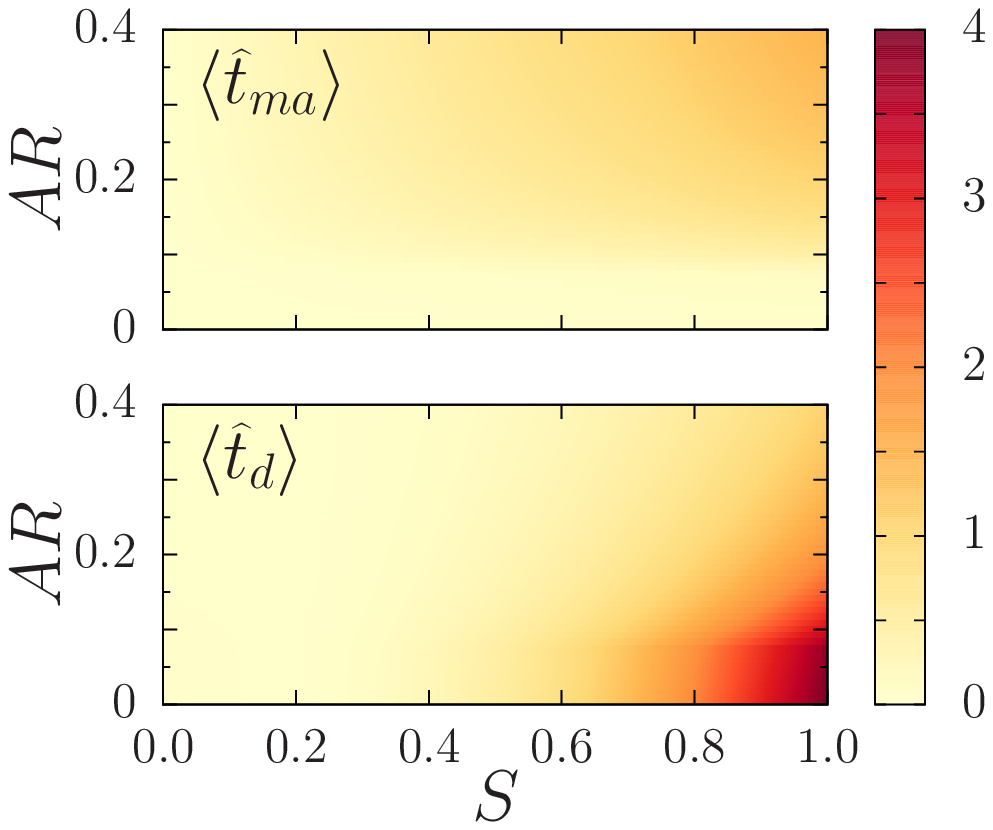}\\
&\\
\hline 
&\\
(c)&(d)\\
\includegraphics[height=0.4\linewidth]{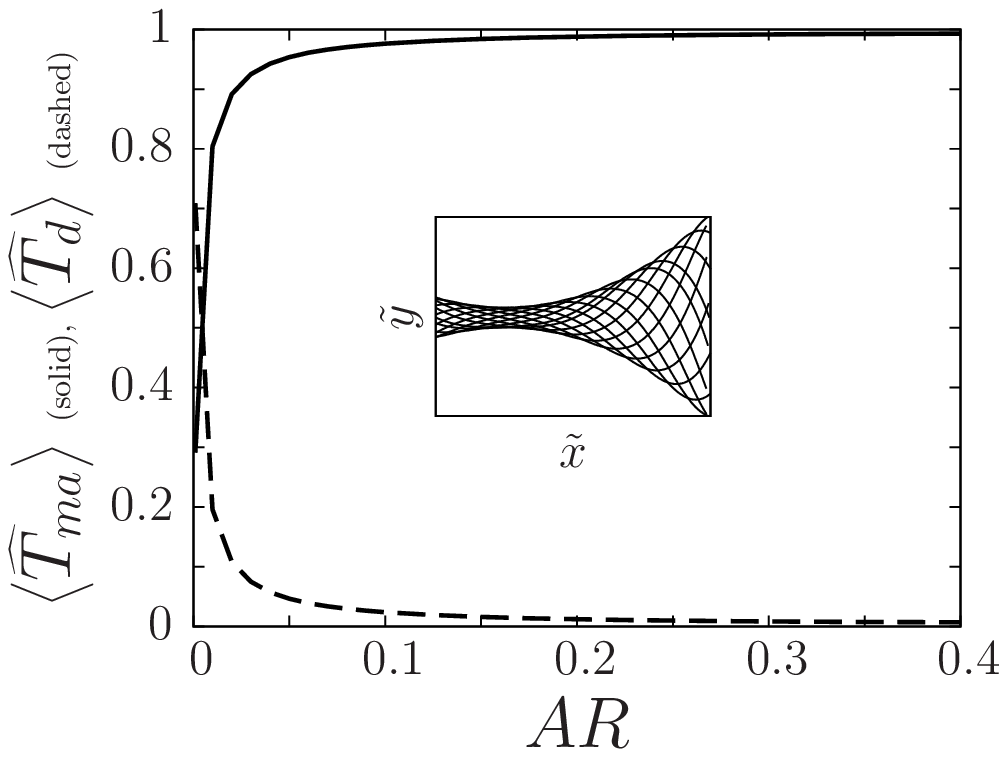}&
\includegraphics[height=0.4\linewidth]{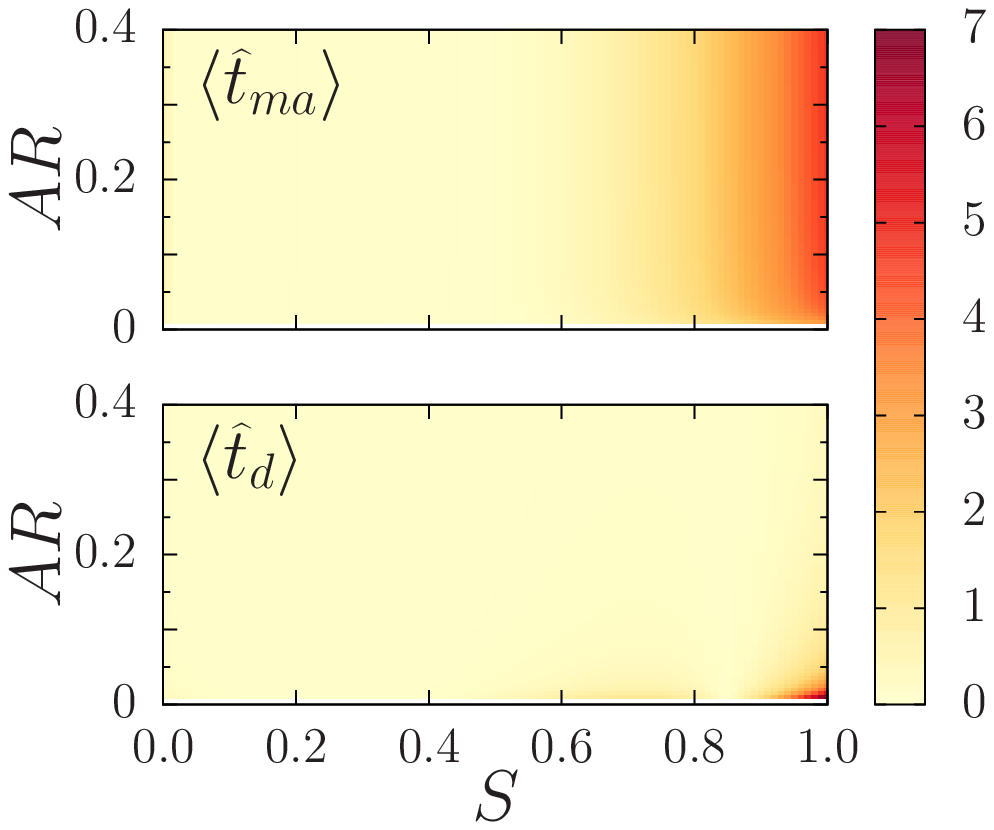}
\end{tabular}
\caption{(a) and (c):  Mean global thrust ratio  as a function of the aspect ratio $AR$ over one oscillating period  for anguilliform and carangiform swimmers respectively. Inset : profiles of the body deformation. (b) and (d) : Normalized mean local thrust ratio as a function of the curvilinear coordinate $S$ over one oscillating period, for anguilliform and carangiform swimmers respectively. \revtres{Where : $<\widehat{t}_{ma}>=<t_{ma}/\int_0^Lt_{ma}ds>$ and $<\widehat{t}_{d}>=<t_{d}/\int_0^Lt_{d}ds>$.}\label{ForceKinemA}}
\end{figure}

Having established the appropriate expressions for $C_d$ and $\mathcal{M}(h)$, the role of the reactive (Eq. \ref{Fmax}) and resistive (Eq. \ref{Fdx}) terms in the dynamical balance that governs the locomotion problem is therefore determined by the swimming kinematics. In the following, we will consider kinematics extracted from real swimmers both for anguiliform and carangiform archetypal species \cite{Gray:1933,Videler:1984,Webb:1984}.
The specific extracted kinematic parameters are the beating amplitude $A$ (as a function of the spinal-cord coordinate $s$), the instantaneous wave speed of the bending wave $v_{\varphi}$ and the instantaneous swimming speed $U$.  These parameters are used to calculate the spatial and time derivatives of $y(s,t)$ which are inserted in the expressions for $t_{ma}$ and $t_d$ (Eqs. \ref{Fmax} and \ref{Fdx}).  
Figure \ref{ForceKinemA} shows contours of the normalized temporal mean of the global added mass and resistive generated thrusts :
\begin{eqnarray*}
&\langle\widehat{T}_{ma}\rangle=\left<\frac{\int_0^Lt_{ma}ds}{\int_0^Lt_{ma}ds+\int_0^Lt_{d}ds} \right>,\\
&\revtres{\langle\widehat{T}_{d}\rangle=\left<\frac{\int_0^Lt_{d}ds}{\int_0^Lt_{ma}ds+\int_0^Lt_{d}ds} \right>},
\end{eqnarray*}
 for two examples of an anguilliform and carangiform swimmer: an eel (taken from the data of Gray's work \cite{Gray:1933}, see inset in Fig.~\ref{ForceKinemA}-a) and a Mackerel (extracted form Videler and Hess \cite{Videler:1984}, see inset in Fig.~\ref{ForceKinemA}-c).
These two characteristic kinematics have been used to analyse the resistive  vs. reactive contributions (that we will hereafter also refer as the drag and added mass contributions) to the thrust production as a function of the aspect ratio of the swimmer. Figure \ref{ForceKinemA}-a shows the ratio of the mean generated global thrusts over one oscillation cycle, for varying aspect ratios for a given slip ratio $U/v_{\varphi}\sim0.55$ (extracted from Gray's work).   By definition, the resistive contribution is independent of $H$, thus giving a single longitudinal distribution for all aspect ratios. The reactive contribution, though, is span dependent and tends to amplify with $AR$ (as $\mathcal{M}(h)$, see Appendix \ref{app_addmass}). As seen in  Fig. \ref{ForceKinemA}-a, both the added mass and drag contributions balance for an aspect ratio $\sim0.13$.  Below this critical value,  the drag effects tend to overcome the added mass contribution. Conversely, added mass effects are dominant for aspect ratios over $0.13$. In the particular case of the example shown in Fig. \ref{ForceKinemA}, kinematics are taken from a swimming butterfish \cite{Gray:1933} with aspect ratio $AR\sim 0.055$.  Figure \ref{ForceKinemA}-b shows the thrust generated by each section of fish along the curvilinear coordinate $s$. It can be seen that both drag and added mass thrust are mostly generated near the end of the animal's body.

In the same manner, results in Fig. \ref{ForceKinemA}-c show the kinematics of a typical carangiform swimmer (mackerel) extracted from Videler and Hess \cite{Videler:1984}.  Here, the slip ratio is $\sim0.81$.  Compared to the anguilliform case, we observe that thrust is almost completely generated by added mass effects in the whole range of physical aspect ratios (between 0.05 and 0.4 \cite{vanWeerden:2014}).  Similar to the anguilliform swimmer, thrust is also mostly produced at the end of the animal's body (Fig. \ref{ForceKinemA}-d).

The simple comparison of these two real cases brings an observation worthy to be underlined: the physical mechanism at the origin of thrust production in inertial swimmers can be very different depending on the driving kinematics (anguilliform vs. carangiform) and on the aspect ratio (long vs. short animals). Especially, the resistive term which is usually associated to low Reynolds number swimmers can be as large or even dominate over the added-mass based reactive mechanisms. It has to be noted that, without being \revdos{explicitly} discussed, this observation has already been reported in recent studies \cite{Eloy:2013}. 

Concerning the swimming kinematics, the slip ratio $U/ v_{\varphi}$ and the amplitude distribution along the undulating body seem to be determinant for the selection of thrust production mechanisms. The remainder of the present work is devoted to studying the \revdos{sensitivity} of our model swimmers to these parameters.

\begin{figure*}[htb!]
\begin{tabular}{l @{} l @{} l}
(a)&(b)&(c)\\
\includegraphics[width=0.33\textwidth]{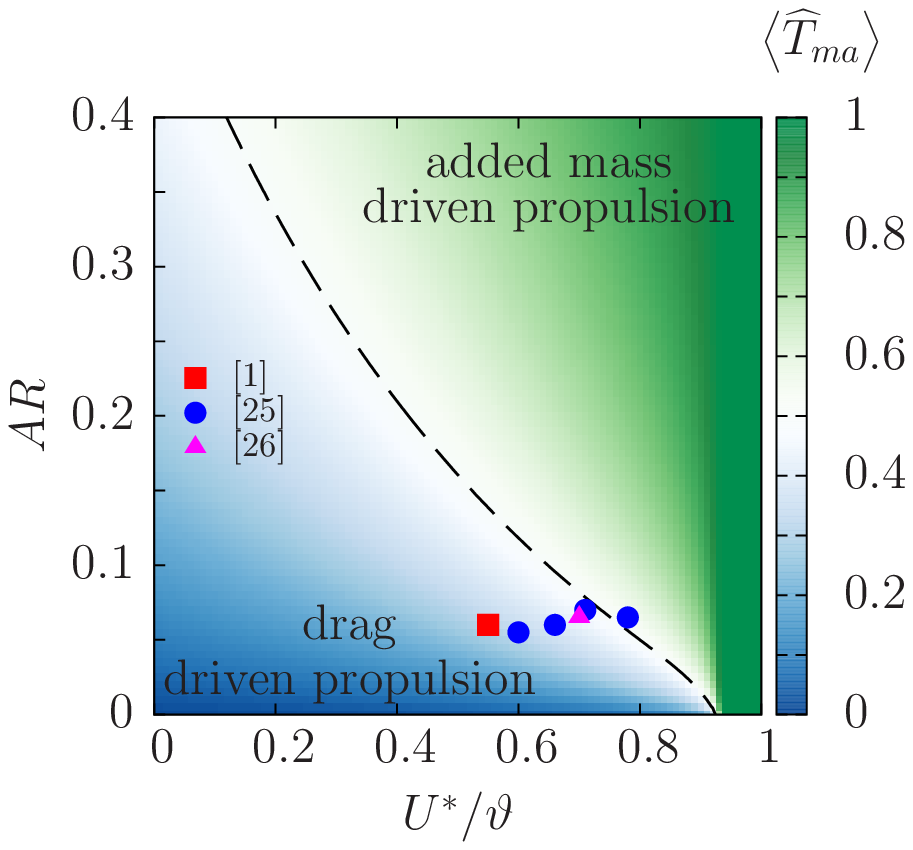}&
\includegraphics[width=0.33\textwidth]{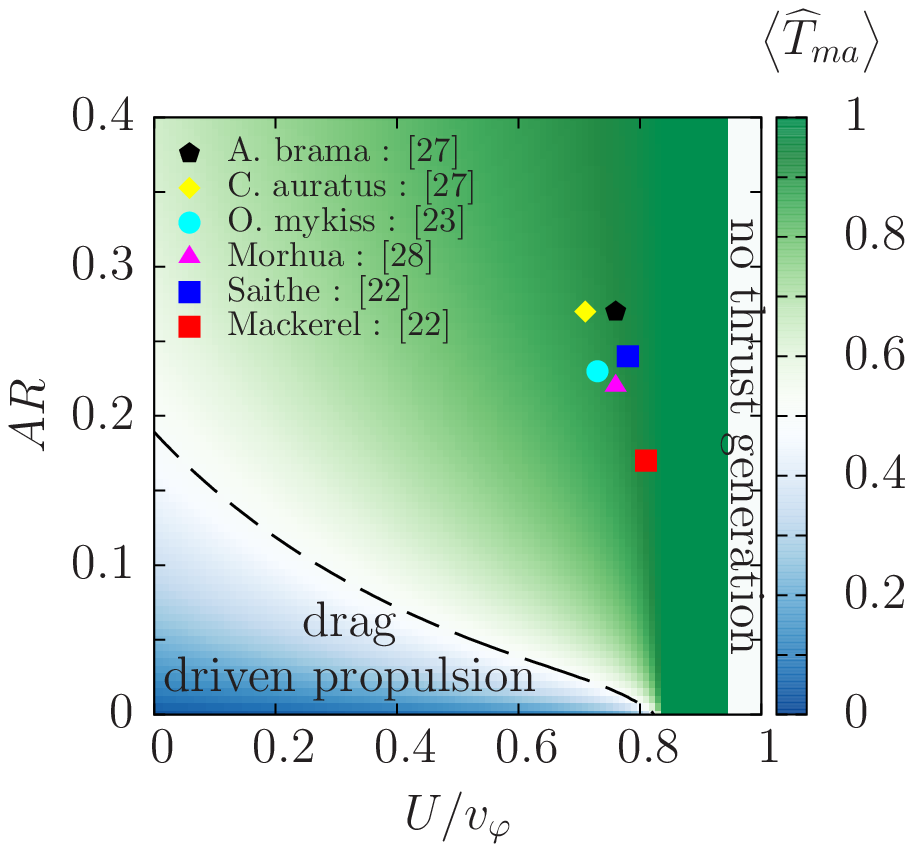}&
\includegraphics[width=0.33\textwidth]{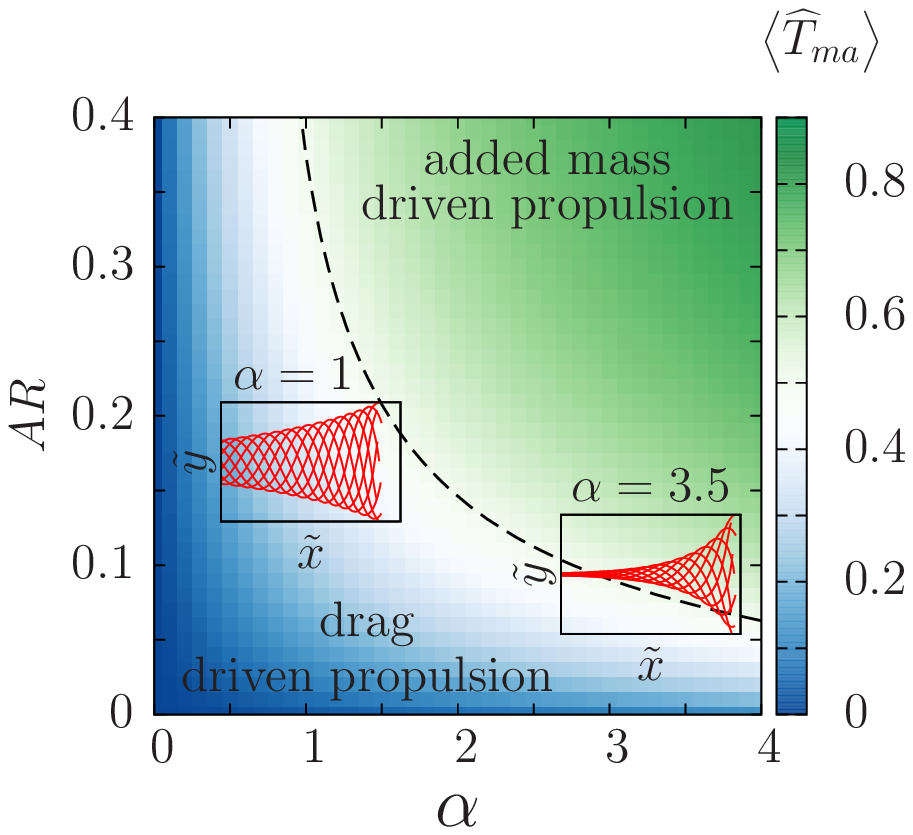}\\
\end{tabular}
\caption{Phase diagrams of drag driven and added mass driven propulsion as a function of the aspect ratio and slip ratio for (a) anguilliform kinematics and (b) carangiform kinematics; and in (c) as a function of the growth rate $\alpha$ of the local body deformation amplitude (see text).  The dashed line represents the $\left<T_{ma}\right>=\left<T_{d}\right>$ in the phase space.  Experimental data is obtained from : Gray, 1933 \cite{Gray:1933}, Tytell, 2004 \cite{Tytell:2004b} and Hess, 1983 \cite{Hess:1983} for anguilliform swimmers, and Bainbridge, 1963 \cite{Bainbridge:1963}, Webb, 1984 \cite{Webb:1984}, Videler, 1978 \cite{Videler:1978} and Videler, 1984 \cite{Videler:1984} for caranguiform swimmers.\label{PDvsU}}
\end{figure*}

First, in order to explore the dependency of the generated global thrust on the aspect and slip ratios, we assume that the deformation profiles remain constant regardless of the swimming and body wave speeds. Fundamentally, kinematics parameters are in some cases interdependent, see for instance discussion in \cite{Tytell:2004a}; however, this hypothesis allows to pinpoint insightful underlying mechanisms. 
Results are shown in Fig. \ref{PDvsU}.  The presented phase diagrams allow to identify drag driven and added mass driven propulsion areas for both anguilliform and carangiform kinematics.  For the case of anguilliform swimmers, small slip ratio values increase the dominance of drag forces in thrust production even for relatively high aspect ratios.  This drag dominance diminishes as the slip ratio increases, until drag propulsion is no more possible for slip ratios $\sim0.92$.  In general, eels and other anguilliform swimmers lie in the rage of aspect ratios $0.05<AR<0.07$, body wavelengths $\lambda/L\sim 0.6$ and slip ratios $0.5<U/v_{\varphi}<0.75$ \cite{Gray:1933,Tytell:2004a,vanWeerden:2014,Borazjani:2009}.  Although the kinematics can vary as a function of the slip ratio, anguilliform swimmers remain in regions where thrust is generated by a comparable contribution between  lateral drag and added mass effects (for high slip ratios).   
In contrast, the carangiform kinematics phase diagram is mostly dominated by added mass thrust production (Fig. \ref{PDvsU}-b).  However, a region of drag dominated propulsion is observed for small slip and small aspect ratios.  Generally carangiform swimmers have aspect ratios around $0.25$, body wavelengths $\lambda/L\sim1$ and have slip ratios much larger than those of anguilliform swimmers, $\sim0.83$ \cite{Videler:1984, Borazjani:2009}.

It is relevant to mention that carangiform swimmers  stay in regions of the kinematic phase space were drag-based thrust production is around zero, avoiding regions were lateral drag will start to produce negative thrust.

Other important difference between the anguilliform and carangiform kinematics presented concerns the amplitude distribution along the body.  While for the anguilliform swimmer the amplitude has almost a linear increment from the head up to the tail, in the carangiform swimmer the lateral displacements of the first part of the body are almost negligible and it is mainly based on the rear half of the moving body. However, both anguilliform and carangiform swimmers in nature adopt varied kinematics (diverse amplitude distributions along the body), differing from the two particular cases presented previously. Several models addressing the global description of anguilliform kinematics have been proposed in the literature to  \cite{Tytell:2004a, Borazjani:2009}.  Following \cite{Tytell:2004a}, we consider that the amplitude distribution of the swimmer is given by :

\begin{equation}
A(s)=A_r e^{\alpha(s-1)},
\label{Amp}
\end{equation}    
where $A_r$ is the amplitude of the displacement at the tail tip of the swimmer and $\alpha$ represents the growth rate of the local amplitude all along the body (i.e. the head to tail amplitude ratio). Fig. \ref{PDvsU}-c shows the regions dominated by either the added mass or the drag contributions to propulsion in an $(\alpha, AR)$ plane and underlines another important effect of the kinematics on the swimming mechanisms: for swimmers using small head to tail amplitude ratio $\alpha <1$ (as sketched in the left insert of Fig. \ref{PDvsU}-c %ARvsAlpha}
), the thrust will be mainly produced by the local drag (i.e. owing to energy dissipation rather than inertia). Increasing $\alpha$ gives more weight to the contribution of added mass mechanisms, which continues to increase as the kinematics tends to that of a carangiform swimmer.\\

Thus, we have shown that introducing a local form drag term to the model describing an idealized inertial swimmer brings a much richer view of the thrust-producing mechanisms than the description commonly used for moderate to large Reynolds number swimmers. As evoked previously, the distinction between resistive or reactive swimmers is usually based on the Reynolds number. For very low Reynolds numbers, swimmers are effectively resistive swimmers just because inertia is missing. In the inertial regime, this distinction is based on both kinematics  (through the ratio $U/ v_{\varphi} $ and $\alpha$) and body geometry ($AR$). For instance, we have seen that slender anguilliform swimmers use a combination of lateral drag and added mass effects in order to generate thrust, with a major resistive contribution for the most slender species. In contrast, typical carangiform swimmers achieve propulsion using mainly the added mass effect, which is predicted by potential flow theories.  

Overall, these results have also an important impact on the design of artificial swimmers. For example, the magnetic swimmers developed by Ramananarivo \textit{et al.} \cite{Ramananarivo:2013}, which  consist of passive flexible filaments (with $AR\sim0.01$) actuated at one end, rely mainly on lateral drag forces to generate thrust, although they swim at moderate Reynolds regimes.  As shown in Figs.  \ref{PDvsU}-a and  \ref{PDvsU}-b, very slender swimmers will indeed rely mostly on drag thrust generation despite their swimming kinematics. Also, due to the nature of their fluid-structure interactions (see for example \cite{Ramananarivo:2014a}), artificial swimmers based on passive flexible structures with imposed pitching or heaving at one edge \cite{Alben:2012,Raspa:2014,Ramananarivo:2013}, tend in general to have wave amplitude distributions with small, or even negative, $\alpha$ values (see Eq. \ref{Amp}).  As shown in Fig. \ref{PDvsU}-c, this can also promote the generation of thrust based mainly on lateral drag effects.  

It is important to note that the conclusions brought with this work are based on the introduction of the local form drag term that accounts for local flow separation all along the body. This contribution, due to tridimensional geometrical effects (the finite size of a fish), cannot be neglected for a correct description of inertial swimmers, but is generally absent in most large Reynolds number swimming studies. We believe that the results raised here may have significant implications not only for the description of swimming in nature but also for future conceptions of inertial artificial swimmers.

\begin{acknowledgments}
We gratefully acknowledge support by EADS Foundation through project ``Fluids and elasticity
in biomimetic propulsion".
\end{acknowledgments}

\appendix
\section{Experimental determination of the added mass coefficient.}
\label{app_addmass}
To estimate the value of the added mass coefficient $\mathcal{M}(h)$, we compare the free oscillations of flexible plates vibrating in air and immersed in a water tank.
\begin{figure}[h]
\begin {tabular} {l}
(a)\\
\includegraphics[width=0.45\textwidth]{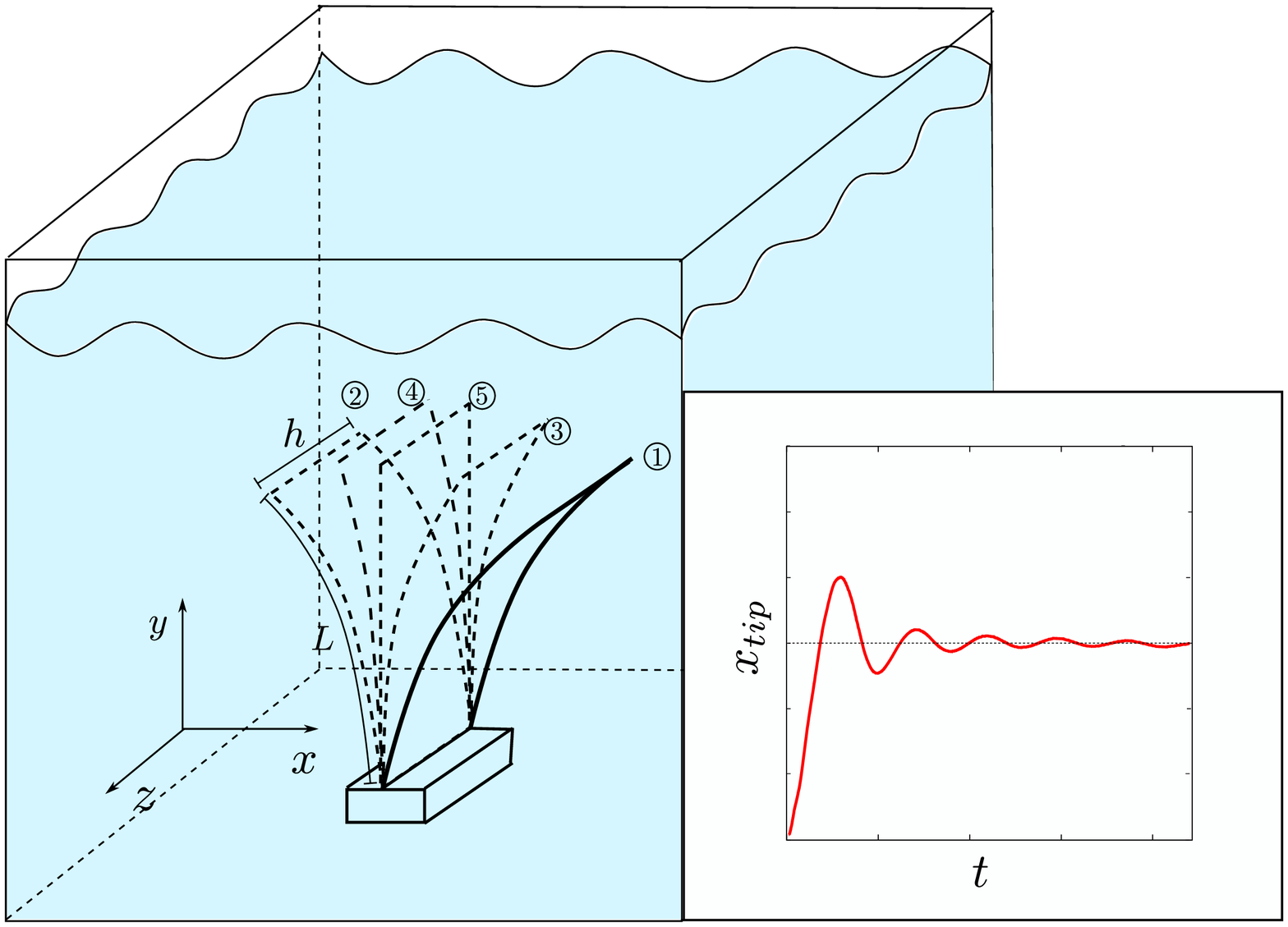}\\
\\
(b)\\
\includegraphics[width=0.35\textwidth]{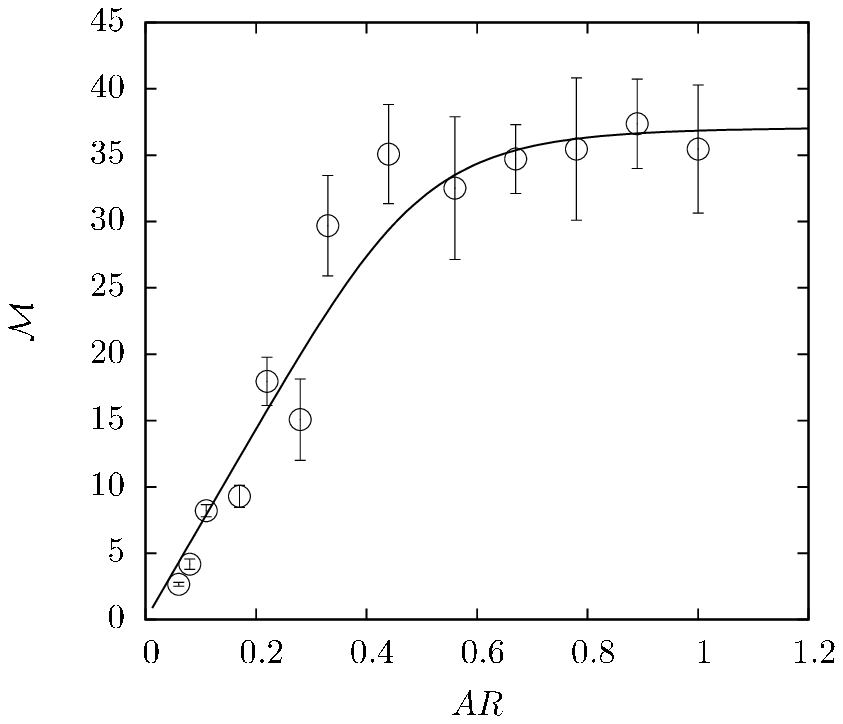}\\
\end{tabular}
\caption{(a) Sketch of the experimental device.  Typical oscillations of the plate's tip position as a function of time.  (b) added mass as a function of the aspect ratio $AR=H/L$ of a rectangular foil.\label{AddMass}}
\end{figure}
The natural oscillating frequency of a cantilevered plate is given by 
\begin{equation}
\omega_n=\frac{\alpha}{L^2}\sqrt{\frac{EI}{\mu + \mathcal{M}}}
\end{equation}
where $E$ is the elastic modulus of the plate, $I$ its moment of inertia, $\mu$ its mass per unit length and $L$ its length.  The non dimensional coefficient $\alpha$ is determined by the mode of deformation of the plate.   

For the experiments performed in air the added mass term is neglected, thus, the added mass coefficient in water can be determined as 

\begin{equation}
 \mathcal{M}=\mu\left(\frac{\omega_{na}^2}{\omega_{nw}^2}-1\right),
\end{equation}
where $\omega_{na}$ and $\omega_{nw}$ are the oscillation frequencies measured in air and water  respectively.

The experimental results for $\mathcal{M}(h)$ are shown in Fig. \ref{AddMass} for aspect ratios ranging from 0.05 to 1.  They correspond to a quadratic dependence of the added mass with $h$ for small aspect ratio (in agreement with elongated body theory \cite{Lighthill:1960}) and a subsequent linear dependance for moderate to large aspect ratios that are consistent with previous results in the literature \cite{Yu:1945,Payne:1981}). A function $\mathcal{M}(h)$ is deduced empirically by fitting the data, following \cite{Payne:1981}, with the function 
\begin{equation}
\mathcal{M}(h)=\frac{\pi \rho h  A_0}{4\left(1+\left(\frac{A_0}{h}\right)^n\right)^{1/n}}\;,
\label{Mh}
\end{equation}

\noindent where $A_0$ is the initial deformation amplitude at the edge of the plate ($x_{tip}(t=0)$ in Fig. \ref{AddMass}-a) and $n=5$.

For the physical range of aspect ratios used in the present work ($0.05<AR<0.4$), the value of the added mass coefficient $\mathcal{M}(h)=\frac{\pi}{4}\rho h$, generally used in elongated body theory, turns to be a good approximation.

%\bibliographystyle{unsrt}
%\bibliography{biblio_swimming}

\begin{thebibliography}{10}

\bibitem{Gray:1933}
J.~Gray.
\newblock Studies in animal locomotion: {I.} the movement of fish with special
  reference to the eel.
\newblock {\em J. Exp. Biol.}, 10(1):88--104, 1933.

\bibitem{Taylor:1952}
G.~I. Taylor.
\newblock Analysis of the swimming of long and narrow animals.
\newblock {\em Proc. Roy. Soc. London. Series A. Mathematical and Physical
  Sciences}, 214(1117):158--183, 1952.

\bibitem{Lighthill:1969}
M.~J. Lighthill.
\newblock Hydromechanics of aquatic animal propulsion.
\newblock {\em Annu. Rev. Fluid Mech.}, 1(1):413--446, 1969.

\bibitem{Triantafyllou:2000}
M.~S. Triantafyllou, G.~S. Triantafyllou, and D.~K.~P. Yue.
\newblock Hydrodynamics of fishlike swimming.
\newblock {\em Annu. Rev. Fluid Mech.}, 32(1):33--53, 2000.

\bibitem{Liao:2007}
J.~C. Liao.
\newblock A review of fish swimming mechanics and behaviour in altered flows.
\newblock {\em Phil. Trans. Roy. Soc. B}, 362(1487):1973, November 2007.

\bibitem{Wu:2011}
T.~Y. Wu.
\newblock {Fish Swimming and Bird/Insect Flight}.
\newblock {\em Annu. Rev. Fluid Mech.}, 43(1):25--58, January 2011.

\bibitem{Hancock:1953}
G.~J. Hancock.
\newblock The self-propulsion of microscopic organisms through liquids.
\newblock {\em Proc. R. Soc. Lond. A}, 217(1128):96--121, 1953.

\bibitem{Gray:1955}
J~Gray and GJ~Hancock.
\newblock The propulsion of sea-urchin spermatozoa.
\newblock {\em J. Exp. Biol.}, 32(4):802, 1955.

\bibitem{Lighthill:1960}
M.~J. Lighthill.
\newblock Note on the swimming of slender fish.
\newblock {\em J. Fluid Mech.}, 9(02):305--317, 1960.

\bibitem{Lighthill:1971}
M.~J. Lighthill.
\newblock Large amplitude elongated-body theory of fish locomotion.
\newblock {\em Proc. R. Soc. Lond. B Biol. Sci.}, 179:125--138, 1971.

\rev{
\bibitem{Wu:1961}
T.~Y. Wu.
\newblock Swimming of a waving plate.
\newblock {\em J. Fluid Mech.}, 10:321--344, 1961.
}

\bibitem{Alben:2012}
S~Alben, C~Witt, TV~Baker, E~Anderson, and G~Lauder.
\newblock {Dynamics of freely swimming flexible foils}.
\newblock {\em Phys. Fluids}, 24:051901, 2012.

\bibitem{Raspa:2014}
V~Raspa, S~Ramananarivo, B~Thiria, and R~Godoy-Diana.
\newblock {Vortex-induced drag and the role of aspect ratio in undulatory
  swimmers}.
\newblock {\em Phys. Fluids}, 26:041701, 2014.

\bibitem{Gazzola:2014}
M.~Gazzola, M.~Argentina, and L.~Mahadevan.
\newblock Scaling macroscopic aquatic locomotion.
\newblock {\em Nature Physics}, 10(10):758--761, September 2014.

\rev{
\bibitem{Gazzola:2015}
M.~Gazzola, M.~Argentina, and L.~Mahadevan.
\newblock Gait and speed selection in slender inertial swimmers.
\newblock {\em Proc. Natl. Acad. Sci.}, 112 (13):3874-3879, 2015.
}

\bibitem{Eloy:2013}
C.~Eloy.
\newblock On the best design for undulatory swimming.
\newblock {\em J. Fluid Mech.}, 717:48--89, February 2013.

\bibitem{Ramananarivo:2013}
S~Ramananarivo, R~Godoy-Diana, and B~Thiria.
\newblock {Passive elastic mechanism to mimic fish-muscle action in
  anguilliform swimming}.
\newblock {\em J. Roy. Soc. Interface}, 10(88):20130667--20130667, 2013.

\rev{
\bibitem{Porez:2014}
M.~Porez, F.~Boyer, and A.~J. Ijspeert.
\newblock Improved lighthill fish swimming model for bio-inspired robots:
  Modeling, computational aspects and experimental comparisons.
\newblock {\em The International Journal of Robotics Research},
  33(10):1322--1341, 2014.
}

\bibitem{Lian:1989}
Q.~X. Lian and Z.~Huang.
\newblock {Starting flow and structures of the starting vortex behind bluff
  bodies with sharp edges}.
\newblock {\em Exp. Fluids}, 8(1-2):95--103, October 1989.

\bibitem{Eloy:2012}
C.~Eloy, N.~Kofman, and L.~Schouveiler.
\newblock {The origin of hysteresis in the flag instability}.
\newblock {\em J. Fluid Mech.}, 691:583--593, January 2012.

\rev{
\bibitem{White:1998}
F.~M. White.
\newblock {\em Fluid mechanics, 4th Edition}.
\newblock McGraw Hill, 1998.
}

\bibitem{Videler:1984}
J.~J. Videler and F.~Hess.
\newblock Fast continuous swimming of two pelagic predators, saithe (pollachius
  virens) and mackerel (scomber scombrus): a kinematic analysis.
\newblock {\em J. Exp. Biol.}, 109(1):209--228, 1984.

\bibitem{Webb:1984}
P.~W. Webb.
\newblock {Body Form, Locomotion and Foraging in Aquatic Vertebrates}.
\newblock {\em Am. Zool.}, 24(1):107--120, 1984.

\bibitem{vanWeerden:2014}
J.~F. van Weerden, D.~A.~P. Reid, and C.~K. Hemelrijk.
\newblock {A meta-analysis of steady undulatory swimming}.
\newblock {\em Fish and Fisheries}, 15(3):397--409, January 2014.

\bibitem{Tytell:2004b}
E.~D. Tytell.
\newblock The hydrodynamics of eel swimming. {II}. effect of swimming speed.
\newblock {\em J. Exp. Biol.}, 207:3265--3279, 2004.

\bibitem{Hess:1983}
F.~Hess.
\newblock Bending moments and muscle power in swimming fish.
\newblock {\em Proc. 8th Australasian Fluid Mechanics Conference. University of
  New Castle, New South Wales}, 2:12A.1--12A.3, 1983.

\bibitem{Bainbridge:1963}
R~Bainbridge.
\newblock Caudal fin and body movement in propulsion of some fish.
\newblock {\em J. Exp. Biol.}, 40(1):23--, 1963.

\bibitem{Videler:1978}
JJ~Videler and CS~Wardle.
\newblock New kinematic data from high-speed cine film recordings of swimming
  cod (gadus-morhua).
\newblock {\em Netherlands Journal Of Zoology}, 28(3-4):465--484, 1978.

\bibitem{Tytell:2004a}
E.~D. Tytell and G.~V. Lauder.
\newblock The hydrodynamics of eel swimming. {I}. wake structure.
\newblock {\em J. Exp. Biol.}, 207:1825--1841, 2004.

\bibitem{Borazjani:2009}
I.~Borazjani and F.~Sotiropoulos.
\newblock Numerical investigation of the hydrodynamics of anguilliform swimming
  in the transitional and inertial flow regimes.
\newblock {\em J. Exp. Biol.}, 212(4):576--592, February 2009.

\bibitem{Ramananarivo:2014a}
S~Ramananarivo, R~Godoy-Diana, and B~Thiria.
\newblock {Propagating waves in bounded elastic media: Transition from standing
  waves to anguilliform kinematics}.
\newblock {\em EPL (Europhysics Letters)}, 105:1--5, March 2014.

\bibitem{Yu:1945}
Y. T.~Yu
\newblock {Virtual Masses of Rectangular Plates and Parallelepipeds in Water}.
\newblock {\em J. Appl. Phys.}, 16:724--729, June 1945.

\bibitem{Payne:1981}
P.R.~Payne
\newblock {The virtual mass of a rectangular flat plate of finite aspect ratio}.
\newblock {\em Ocean Eng.}, 8:541--545, 1981.

\end{thebibliography}

\end{document}